# Evaluating security vulnerabilities in web-based Applications using Static Analysis

Chinwuba Christian Nnaemeka
Department of Computing & informatics
Bournemouth University
Poole, United Kingdom
S5433891@bournemouth.ac.uk

Osejobe Ehichoya
Department of Computing & informatics
Bournemouth University
Poole, United Kingdom
S5433891@bournemouth.ac.uk

*Abstract*—Web services are becoming business-critical components, often deployed with critical software bugs that can be maliciously explored. Web vulnerability scanners allow the detection of security vulnerabilities in web services by stressing the service from the point of view of an attacker. However, research and practice show that different scanners perform differently in vulnerability detection. This paper presents a qualitative evaluation of security vulnerabilities found in web applications. Some well-known vulnerability scanners have been used to identify security flaws in web service implementations. Many vulnerabilities have been observed, which confirms that many services are deployed without proper security testing. Additionally, having reviewed and considered several articles, the differences in the vulnerabilities detected and the high number of false positives (35% and 40% in two cases) and low coverage (less than 20% for two scanners) observed highlight the limitations of web vulnerability scanners in detecting security vulnerabilities in web services.

Furthermore, this work will discuss the static analysis approach for discovering security vulnerabilities in web applications and complimenting it with proven research findings or solutions. These vulnerabilities include broken access control, cross-site scripting, SQL injections, buffer overflow, unrestricted file upload, broken authentications, etc. Web applications are becoming mission-essential components for businesses, potentially risking having several software vulnerabilities that hackers can exploit maliciously. A few Vulnerability scanners have been used to detect security weaknesses in web service applications, and many vulnerabilities have been discovered, thus confirming that many online apps are launched without sufficient security testing. The static analysis technique considered in this work helps detect security flaws. However, it has an important limitation of indicating high false positives.

## I. Introduction

The importance of web-based application security has increased since it handles sensitive data that, if hacked, may cost the company millions of dollars [19]. Vulnerabilities in web applications significantly impact application security and users' risk of being attacked by hackers exploiting flaws in the source code of web applications. How to detect security bugs efficiently in software systems is a growing concern, and it is vital to secure these applications against hackers [19]. In 2002, the computer security institute and the FBI performed a computer crime and security assessment, which indicated that more than half of all databases experienced at least one security breach, with an average loss of about $4 million [10]. In a survey by Alqaradaghi et al., 2021 about 75% of all attacks on web servers target web-based applications, and firewalls cannot defend against them because they rely solely on HTTP traffic, which is typically allowed to pass through them unimpeded. Attention has been focused on network-level attacks like port scanning. As a result, hackers frequently gain access to Web apps directly.

Developers' lack of proper understanding of secure coding is a primary cause of web application insecurity, resulting in flaws. Many ways of detecting source code vulnerabilities have been investigated, with some extant approaches falling into two categories: dynamic and static analysis. Dynamic analysis approaches, such as software testing, examine how an application program runs but only guarantee 100% coverage. In contrast, the static analysis examines the application's source code with many false positives but achieves 100% testing coverage [22].

According to a recent web security analysis and research, cross-site scripting (XSS) is the most susceptible web application vulnerability [23]. They are inserted into the web applications' source code without encrypting or verifying XSS scripts. Hackers exploit them to steal sensitive data, cookies, and web sessions. XSS vulnerabilities are produced when malicious scripts are hosted on a website or when a malicious URL lures a user. This vulnerability affects web applications and is a known concern [20].

This paper examines a technique for discovering security vulnerabilities in web-based applications using static analysis. The methods entail studying the web application source code for input validation defects and putting solutions into the same principle to repair these flaws. This technique immediately adds to web application security by reducing vulnerabilities and indirectly allowing developers to identify the problems.

The paper is organised into five main sections, with section one having four sub-sections. The main sections include the introduction, methodology, research findings, conclusion, and recommendation. Main section one has four sub-sections: background, literature review, problem statement and research questions.

*A. Background*

- *Static Analysis*

Static analysis of web applications is indispensable to any software tool, integrated development environment, or system that requires compile-time information about the semantics of programs. With the emergence of modern programming languages, static analysis of applications consisting of both recursive data and dynamic storage has become a field of active research. Programming mistakes introduce vulnerabilities in program source code that needs to be fixed. The longer a vulnerability lies dormant, the more expensive it can be to fix. Static analysis tools aim to identify common coding mistakes before an application is deployed automatically. Static analysis extracts semantic information about a program at compile time [31]. It verifies the program source code statically without attempting to execute the code. Manual code auditing is a form of static analysis which is time-consuming and requires the code auditors first to understand what security flaws appear like to check the code thoroughly. Static analysis tools are quicker than manual code audits, as they regularly analyse programs.

Additionally, because they are designed to capture security knowledge, they do not need the tool operator to have the exact level of security experience as a human auditor [32]. Static analysis will only solve some of your security issues. For instance, it scans the code for a predefined collection of patterns or criteria. To eliminate vulnerabilities, experienced developers must still design a program correctly. Static analysis techniques can detect bugs in the essential details, but they cannot evaluate the design. The result of a static analysis tool still needs human assessment [33].

- *Vulnerability*

Web applications are accepted in today's business environment and are used in the business's day-to-day activities. Several companies have launched Web applications, and their use has recently surged. As web-based applications become more critical business elements, they are frequently deployed with significant software vulnerabilities that can be exploited illegally. Vulnerabilities are defects or weaknesses in a system's architecture, development, and operation that might be manipulated to break the system's security procedures or functionality [36]. Any vulnerability or hole in a web application can be used to obtain unauthorised access to, harm, or corrupt the information system. Web application vulnerabilities are embedded in web application codes. They are unaffected by the technology used to develop the application, the safety of the Web application, or the back-end system. Developers should follow proper coding practices, thoroughly evaluate the code for security vulnerabilities, run penetration testing, and employ code vulnerability checkers to prevent vulnerabilities. Technically, the cost to fix security flaws discovered later in the software development cycle is higher than security flaws found earlier [24]; developers must make every effort to identify problems as soon as feasible. Code audits (code reviews), static analysis, dynamic analysis, and security testing are methods for identifying vulnerabilities in online applications [35]. White-box and black-box testing are the two primary techniques for vulnerability testing in web applications. Black-box testing does not directly examine the source code of the program, unlike white-box testing, which does so to find defects or vulnerable lines of code. Static analysis is an example of white-box testing [34].

*B. Literature Review*

Web application vulnerability predicting frameworks were built using historical data that showed the proposed known vulnerability data along with static properties to anticipate the XSS and SQLI flaws [6]. A set of static code attributes was intended to represent these code patterns. They enhanced their work process by developing a strategy for building construction predictors using machine learning language [7]. A prototype program called PhpMinerl was created to collect data and evaluate its methodology on different open-source web applications. The results showed 11% false positives in detecting SQLI vulnerability and 6% in detecting XSS in web applications. The developers were trained using the available vulnerability exposure dynamic and static analysis data. The static analysis primarily evaluates the program source code without executing the regulations, whereas dynamic analysis examines how this application works by code execution and validating its functionalities. Pixy was the first static analysis tool to discover the XSS vulnerabilities in PHP source code in 2006; the report from this research suggests that using the right static analysis tool achieves a successful outcome of approximately 72% predicting XSS vulnerability in web applications and the result of static code analysis used reported a false positive rate of around 9%. Nevertheless, XSS vulnerability in web applications persists due to analytical limitations, such as the false positive rate on the analysis's findings [25].

The reliability of security scanners in discovering vulnerabilities vary, and it is an excellent tool for finding web application flaws introduced in the source code during development. Web security scanners are one method that frequently characterises the effectiveness of various scanners in identifying vulnerabilities in online applications [27]. Web security scanning tools are tested to determine their strengths and limitations regarding vulnerability assessment coverage and false positives. The objective was to investigate the reliability of security scanners and identify effective forms of web application vulnerabilities. Three commercial scanning technologies were assessed, and the results revealed that overall coverage is inadequate, with many false positives [13].

Nonetheless, the analysis was limited to a particular family of software, mainly web-based apps developed in PHP. The findings cannot be generalised because many services examined were launched without sufficient security testing [4]. The web scanner tools are divided into enterprise-level and free, open-source tools; the enterprise-level device has been evaluated as more accurate and precise due to the implementation of extra innovations. Furthermore, in the context of describing techniques for mitigating SQL injection attacks, enterprise tools provided a transparent, solidly automated SQL injection analysis tool based on a syntax algorithm [26]. A significant finding was that different scanners discover different vulnerabilities, indicating that one scanner's coverage is far from flawless. Web scanners have a high false-positive rate and low range, exposing their limits in finding vulnerabilities in web applications [13].

Numerous studies have examined techniques for comprehensively evaluating various web threats, including SQL injection, XSS, and other vulnerability mitigation and detection techniques, to understand better the general engineering fields connected to web security threats [18]. Static analysis tools can provide a reliable warning to some extent, according to Walden and Doyle's research [16], which found a strong link between Fortify SCA tool alerts and NVD vulnerabilities. Zheng et al. [17] showed how static analysis is a crucial technique for identifying flaws that have the potential to lead to security vulnerabilities based on an enterprise-scale investigation. In comparing the value of manual code review with static analysis (black-box testing) for online applications, Finifter and Wagner [15] found that the two are complementary and that manual analysis revealed more errors but consumed a lot more effort and needed experts to examine the application code base. They argued that no single technique could find every vulnerability in a web application. Their research revealed that relatively rare vulnerabilities are discovered using a variety of methodologies, with automated penetration testing being the most efficient in terms of time and static analysis coming in second. Research, parameter fiddling, SQL injection, and cross-site scripting attacks contribute to more than a quarter of all identified Web application vulnerabilities [45]; the attacks listed above are made possible by user input that has not been adequately validated. Coding auditing can detect these attacks, and code reviews discover issues before launching a program. Code reviews are one of the most effective defence measures [17], but they are time-consuming and expensive. Thus, they are only used sparingly. Many programmers lack the security expertise required for security audits, which drives up the cost of security assessments. Since security issues are regularly introduced as they are being fixed, double audits (examining the code twice) are strongly encouraged.

*C. Problem Statement*

As web technologies advance and users shift away from traditional desktop applications, the adoption of web-based applications has surged. Among the professional developers who design web applications are a few amateurs with limited knowledge of web application security who create vulnerable applications. These security vulnerabilities allow attacks to gain unauthorised access to the web application. The most prevalent cause of web applications is unchecked input parameters in the source code, which is a typical development error [28]. Hackers employ two approaches to attack uncontrolled input parameters: they inject malicious code into web applications and then use the code to manipulate the application. In 2013, the (OWASP) Open Web Application Security Project identified the most severe web application security vulnerabilities [20].

1. SQL Injection Attacks
2. XSS-Cross Site Scripting
3. Broken Authentications and Session Management
4. Sensitive Data Exposure
5. Security Misconfigurations

*D. Research Question*

- How often does vulnerability appear in a web-based application?
- What level of breach or impact do these vulnerabilities cause?
- What are the existing static analysis techniques, and how best can they be improved to optimise performance?

II. METHODOLOGY

This study used a qualitative research technique to critically analyse existing web application static analysis approaches or strategies. The investigators conducted a systematic review of academic papers in current peer-reviewed journals to assess the secondary literature on the research. Systematic reviews summarise what has been written and discovered about a research topic objectively. This is especially useful when several articles on an extensive study topic each focus on a different aspect of the field. The investigators will conduct their research using the databases MySearch, Google Scholar, IEEE, Scopus, ScienceDirect, and Web of Science. These databases are multidisciplinary, well-established research platforms regularly updated and feature a wide range of peer-reviewed publications. These databases were chosen to include all relevant papers. The researcher's evaluation will be restricted to peer-reviewed literature. Peer-reviewed journal articles are expected to contain high-impact research on Web-based application static analysis. The researchers established a cut-off year for the review to ensure that the data gathered was relevant, and they investigated its impact on the field.

In this research, we have considered a wide range of literature from 2010 - 2022. This range of literature will present comprehensive state-of-the-art research conducted in this field

*A. Exclusion and Inclusion Metric*

The systematic search strategy included phrases such as "static analysis," "SQL Injection," and "cross-site scripting." The systematic review concentrated on how much research had been conducted on web-based application static analysis. Other keyword phrases, such as "web-based application vulnerability" or "web applications attack," were used in the search, but the results were limited to "Static Analysis" only. As a result, searches such as "web application static analysis" were carried out. The search terms were intended to draw attention to the subject under investigation so that relevant evidence could be discovered [3]. Below, in Fig 1, is the Prisma flowchart describes how our research was conducted and the databases where relevant research articles were found.

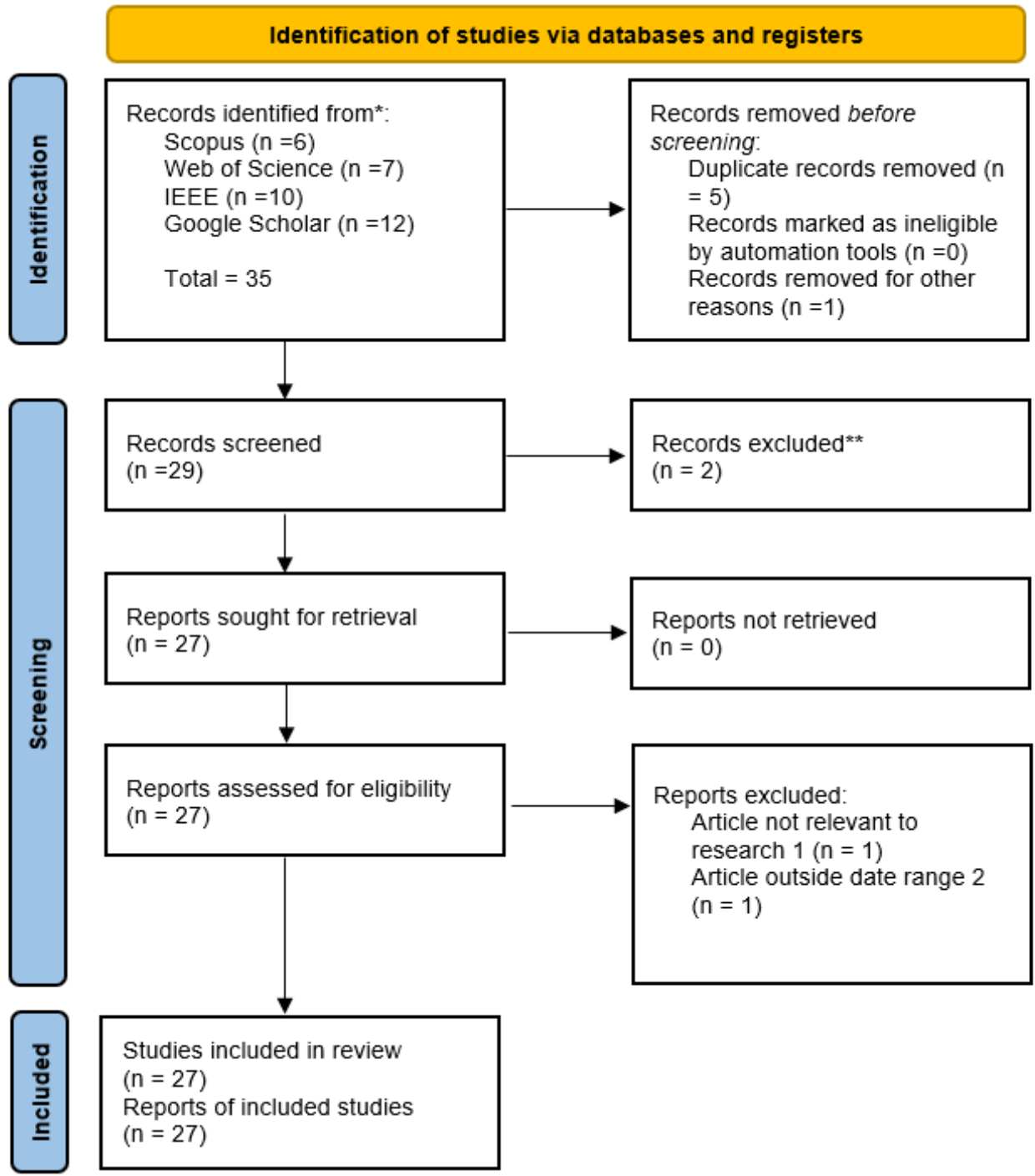

*Fig 1. Prisma flow diagram*

## III. RESEARCH FINDINGS

The research questions addressed in the study were as follows: (1) How frequently does vulnerability appear in a web-based application? (2) How serious is the breach or impact caused by these vulnerabilities? (3) What are the current static analysis techniques, and how can they be enhanced to improve performance? The data was compiled through a thorough review of peer-reviewed articles. Below are some of the prominent findings discovered in our research.

### A. Impact of Web-based Application Attacks

According to the annual global security report 2018, which analysed billions of security events, all tested applications have at least one vulnerability and an average of 11 failures. Web attacks appear to be becoming more specific, frequent, and sophisticated [39]. A successful web-based attack can significantly impact websites, web applications, reputation, and customer relationships. It defaces the websites, compromises user accounts, runs malicious code on web pages, etc., potentially compromising the user's device. It stems from poorly developed web applications' source codes which are not adequately checked. According to a TechJury report, 30,000 websites are hacked daily [37].

Frequently, hackers target financial, healthcare, and retail organisations, and if cybercriminals cannot breach an organisation's security infrastructure, they may attempt to gain access to the corporate website. Similarly, software vulnerability and third-party integrations such as extension usage are also ways attackers can accomplish this. Some plugins are responsible for 98% of the vulnerabilities in content management systems, such as WordPress, which hosts over 35% of all websites on the internet. As a result, numerous security plugins are available to protect the vulnerable.

### B. Attack Vectors and Enablers

Web applications can be attacked for various reasons, including system flaws caused by incorrect coding, misconfigured web servers, application design flaws, or failure to validate forms. SQL injection (SQLI), cross-site scripting (XSS), remote code execution (RCE), and file inclusion (FI) are among the most common and severe web application vulnerabilities threatening the privacy and security of both clients and applications today, according to OWASP's Top 10 Project [7]. These flaws and vulnerabilities enable attackers to access databases containing sensitive information. Web applications are an easy target for attackers because they must always be available to customers. According to ENISA [4] threat report, there is a general perception that web application attacks are diverse. However, other data from security research suggests that most web application attacks are limited to SQL injection or Local file inclusion. Another report indicates that SQL injection, directory traversal, XSS, broken authentication and session management are at the top of the attack vectors used in Web application type of attack. SONICWALL also reported a similar trend for the top web application attacks for 2019. On the list, SQLi, directory traversal, XSS, broken authentication and session management were on the top.

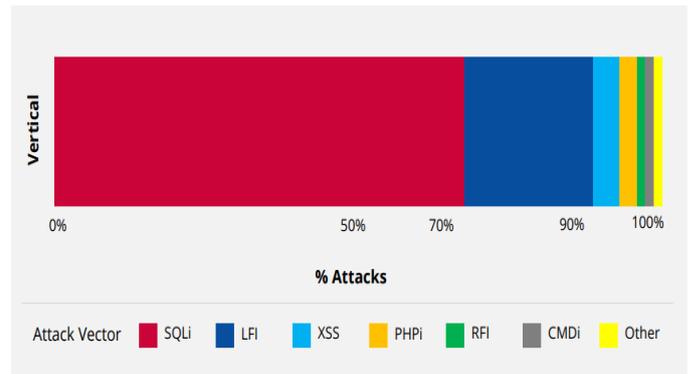

*Fig 2. Threat landscape [4]*

The findings also indicate that the Cross-Site Scripting (XSS) vulnerability is most common in web applications. This vulnerability can result in violations for the user or the site. Many tools and methods focus on finding this vulnerability in PHP source code. Nonetheless, identifying XSS vulnerabilities in PHP web applications remains a challenge for the time being. Most previous tools and approaches relied on static analysis to detect XSS vulnerabilities. This is due to its ability to achieve nearly 100% code coverage and observe all programme paths. Furthermore, recent research has found that static analysis is superior to other approaches for detecting this vulnerability. Combining static analysis with other algorithms (genetic algorithms, pattern matching, and machine learning) improved detection results and reduced static analysis run time [6].

### C. Prevention Mechanism

Various research articles have investigated practical and comprehensive approaches to vulnerability in Web-based applications. According to [5,] existing mechanisms for dealing with Web application threats can be divided into client-side and server-side solutions. An application-level firewall protects against cross-site scripting (XSS) attacks that try to steal a user's credentials. Server-side solutions have the advantage of discovering a broader range of vulnerabilities.

Pixel, [5] according to one of the peer-reviewed articles, was the first open-source tool for statically detecting XSS vulnerabilities in PHP 4 code using data flow analysis. PHP was chosen as the target language because it is widely used for developing Web applications, and many security advisories mention PHP programs. Although the peer-reviewed article considered a prototype designed to detect XSS flaws, it can also be used to detect other taint-style vulnerabilities such as SQL injection or command injection. Any significant type of vulnerability (for example, cross-site scripting or SQL injection) can be considered an example of this general class of taint-style vulnerabilities [5].

Pixy was tested in this manner using six popular open-source PHP programs, and the test result returned accurate results [5].

The extensive research conducted in [2], as shown in Fig 3 below, presents an approach for discovering and correcting vulnerabilities in web applications and a tool that implements the policy for PHP programmes and input validation vulnerabilities. The method and device look for vulnerabilities by combining two techniques: static source code analysis and data mining. The top three machine learning classifiers are used to identify false positives, and an induction rule classifier is used to justify their presence. Static analysis tools assist lowers the price of application maintenance via early detection and avoidance of problems in web applications, making static analysis tools an essential framework in defending against web application attacks. Adjustments to source code can be quickly checked to increase code security with the introduction of static analysis tools into the CI/CD pipelines [24]. These technologies can help web-based applications by detecting problems in the model, minimising security failure mechanisms, and highlighting areas for development. Code analysis tools are mainly used to detect programming language problems and code syntax incompatibilities [29]. These tools are promising for improving the robustness of web-based application security. Nevertheless, it suffers a significant setback from a high false positive rate when examining source codes. The adverse effects of this high false positive rate have led to a lacklustre uptake of these tools. False positives in code analysis are thought to be a complex problem [30]. Validation must balance false positives and negatives (missing defects) (no defect present). To reduce false positives, contextualised error reporting, conflicting terminology and issue prioritizations must all be considered by static analysis tools running on web-based applications.

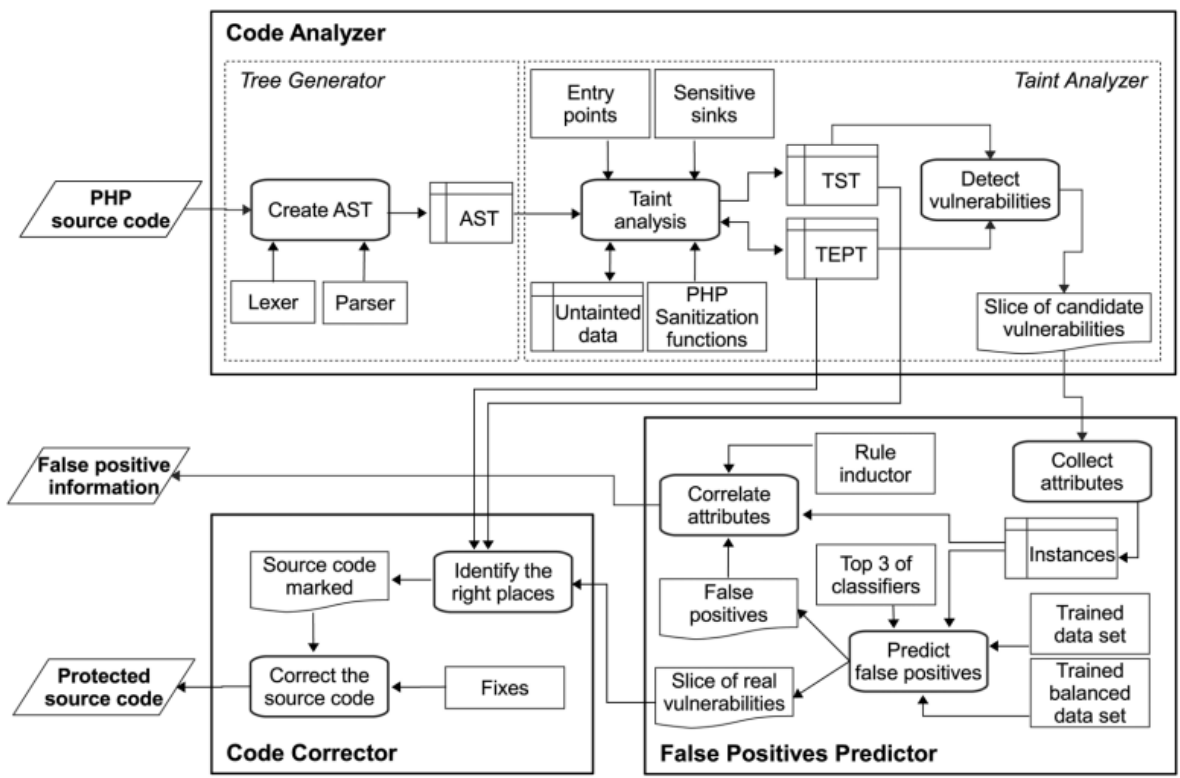

*Fig 3. Main modules and data structures. [2]*

*D. Widespread Awareness Campaign*

Even though much research is being done to mitigate vulnerabilities in web applications, raising awareness about this issue is still critical. Both web application developers and users must be aware of the gravity of web vulnerabilities and what they can do to mitigate their impact on web application security. Security requirements should be integrated into web application development at all stages of the software development lifecycle.

Many web applications are created quickly, and security is an afterthought. It is critical for web application developers to understand not only the negative impact of XSS and other vulnerabilities but also to be able to mitigate them; by so doing, it would assist developers in addressing web vulnerabilities when developing web applications. Similarly, web application users should be warned and given best practice guidelines when visiting web applications online, especially when sensitive information is shared. Some banking applications, for example, warn visitors about security risks. Furthermore, users must exercise caution when clicking links that may direct them to an insecure site where they may become victims of hackers [38].

## IV. CONCLUSION

The principal objective of this qualitative study was to investigate static analysis as it relates to web-based applications and its relative impact on a wide scale. A comprehensive literature study and peer-reviewed journal article was conducted during the data-gathering phase. Three themes emerged from the data study: the impact of Web-based Application attacks, attack Vectors and Enablers, and Prevention Mechanisms. This study's findings were used to address the research questions. Relevant stakeholders can use these findings to improve web application security. Web applications have become a popular and widely used medium of interaction in our daily lives. Simultaneously, vulnerabilities that endanger users' data are discovered regularly. Manual security audits aimed at these vulnerabilities are time-consuming, expensive, and prone to error. Therefore, SDLC stakeholders need to be aware that hacking techniques are constantly changing with the advancement of technology, and there are always new ways to steal information from businesses. Thus, protecting web systems may reduce security risks, increase customer confidence, and improve the economy's health.

## V. RECOMMENDATION

This study uses various strategies to show how SMEs protect sensitive firm data from cyber threats. The offered approaches are action plans for industry small business entrepreneurs or MSMEs. As the first guideline, small business owners should establish a company strategy that engages in active cybersecurity actions. Such a strategy should include policies and methods to safeguard corporate and consumer data from cyber threats. The second advice is for small business owners to gradually link their business operations to cybersecurity rules to develop a unified security strategy across their organisation. The final recommendation is for small business owners to build an adequate plan addressing preparation, data privacy, and data breach response in case of a breach, which can help lessen the impacts of data breaches while preserving personal company data.

However, as considered in this work, static analysis of detecting web application vulnerability was thoroughly expanded. The main disadvantage of static analysis is the high rate of false positives in the results. False positives are results seen as vulnerable paths but not weak. Another disadvantage of static analysis approaches is their dependence on a particular framework or language. For example, a static analysis tool designed for PHP cannot be used for Ruby on Rails without extensive engineering work. These tools are known to be inextricably linked to both language and framework features. Considering this shortcoming, it would be ideally suitable to combine static, dynamic and hybrid analysis [24].